\documentclass[aps,prb,twocolumn,showpacs,lettersize]{revtex4-1}
\usepackage{amsmath}
\usepackage{amsthm}
\usepackage{amssymb}
\usepackage{wasysym}
\usepackage{bm}
\usepackage{natbib}
\usepackage{amsbsy}
\usepackage{graphicx}
\usepackage{color}
\usepackage{MnSymbol}
\usepackage[colorlinks=true,linkcolor=red, citecolor=blue]{hyperref}

\begin{document}


\title{Reply to `Comment on ``$Z_2$-slave-spin theory for strongly correlated fermions'''}
\author{Andreas R\"uegg}
\affiliation{Department of Physics, University of California, Berkeley, Berkeley Ca 94700, USA}
\author{Sebastian D. Huber}
\affiliation{Department of Condensed Matter Physics, The Weizmann Institute of Science, 76100 Rehovot, Israel}
\author{Manfred Sigrist}
\affiliation{Theoretische Physik, ETH Z\"urich, CH-8093 Z\"urich, Switzerland}
\date{\today}
\begin{abstract}
We show that the physical subspace in the $Z_2$-slave-spin theory is conserved under the time evolution of the system. Thus, when restricted to the physical subspace, this representation gives a complete and consistent description of the original problem. In addition, we review two known examples from the existing literature in which the projection onto the physical subspace can be relaxed: (i) the non-interacting limit in any dimension at half filling and (ii) the interacting model in the infinite dimensional limit at half filling. In both cases, physical observables are correctly obtained without explicit treatment of the constraints which define the physical subspace. In these examples, correct results are obtained, despite the fact that unphysical states enter the solution.
\end{abstract}
\maketitle


A defining property of so-called ``slave-particle" constructions is that the physical degree of freedom is represented by auxiliary degrees of freedom in an enlarged Hilbert space. In the $Z_2$-slave-spin representation introduced in Refs.~\onlinecite{Huber:2009,Ruegg:2010b}, the local electron operator $c_{i\sigma}$ is represented by an auxiliary fermion $f_{i\sigma}$ and a spin-1/2 slave-spin $\vec{I}_i=(I^x_i,I^y_i,I^z_i)$ as $c_{i\sigma}=2I_i^xf_{i\sigma}$. The physical states $|{\rm phys}\rangle$ form a subspace in the enlarged Hilbert space and are singled out by the local constraints
\begin{equation}
A_i|{\rm phys}\rangle=0, \quad A_i=I_i^z+1/2-(n_i-1)^2,
\label{eq:Ai}
\end{equation}
where $n_i=f_{i\uparrow}^{\dag}f_{i\uparrow}+f_{i\downarrow}^{\dag}f_{i\downarrow}$ denotes the local $f$-fermion density and $i=1,\dots N_s$ runs over all the lattice sites. In terms of these new variables, the original Hubbard model can be written as
\begin{equation}
H_S=-4t\sum_{\langle i,j\rangle\sigma}\left(I_i^xI_j^xf_{i\sigma}^{\dag}f_{j\sigma}+{\rm h.c.}\right)+\frac{U}{4}\sum_i(1-2I_i^z).
\label{eq:Hs}
\end{equation}
In Refs.~\onlinecite{Huber:2009,Ruegg:2010b} we argued that the original problem is equivalent to the slave-spin problem Eq.~\eqref{eq:Hs} acting on the physical Hilbert space defined by Eq.~\eqref{eq:Ai}. Explicitly, the projection operator onto the physical subspace was given in Ref.~\onlinecite{Ruegg:2010b} as:
\begin{equation}
\mathcal{P}=\prod_i(1-Q_i),\quad Q_i=A_i^2.
\end{equation}
The operator $Q_i=A_i^2$ fulfills $[H_S,Q_i]=0$ in the extended Hilbert space but does {\it not} provide an additional constraint on the physical Hilbert space. This follows immediately from the fact that Eq.~\eqref{eq:Ai} {\it implies} $Q_i|{\rm phys}\rangle=0$. Nevertheless, because $Q_i$ has eigenvalues 0 and 1 it is used to construct the projector.

In order that this projective scheme can be implemented consistently, the physical subspace needs to be invariant under the time evolution generated by $H_S$. This is indeed the case: one can verify that
\begin{equation}
B_l:=[A_l,H_S]=\left[8t\sum_{i\sigma}I_i^xI_l^x\left(f_{i\sigma}^{\dag}f_{l\sigma}+f_{l\sigma}^{\dag}f_{i\sigma}\right)\right] A_l
\label{eq:Bl}
\end{equation}
where the sum runs over all sites $i$ connected to site $l$ by the hopping amplitudes. Clearly, $B_l\neq 0$ in the extended Hilbert space but for physical states, it follows that
\begin{equation}
B_l|{\rm phys}\rangle=0,
\label{eq:B}
\end{equation}
i.e., the physical subspace as defined by Eq.~\eqref{eq:Ai} is conserved under the evolution generated by $H_S$. The slave-particle representation given by Eq.~\eqref{eq:Hs} acting on the physical Hilbert space therefore indeed gives a complete and consistent description of the original problem. This is the view-point we have adopted in our original work.\cite{Ruegg:2010b} A more complete discussion of constraint quantum systems, as initiated by Dirac, can be found in the textbooks, see e.g.~Refs.~\onlinecite{Dirac:2001,Henneaux:1992}.

 
In approximative treatments of the slave-spin problem, the restriction to the physical subspace is no longer guaranteed. In particular, 
if we replace the condition Eq.~\eqref{eq:Ai} by the weaker condition $\langle A_i\rangle=0$ for states in the extended Hilbert space, unphysical states are no longer excluded from the theory. The objection formulated by the authors of Ref.~\onlinecite{Ferraz:2013} is built on this observation but is more general: they seem to be concerned by the fact that the mean-field treatment does not correctly enforce the projection onto the physical subspace. 

Here we would like to clarify a potential source of confusion. In our original article,\cite{Ruegg:2010b} we adopted the projective view-point of the slave-spin construction, as reviewed above. In Sec.~VI we also discussed a mean-field treatment which enforces $\langle A_i\rangle=0$ on average by the Lagrange multiplier method. We stress that this approximation (with a static Lagrange multiplier) was not meant as an alternative to the projective scheme -- to obtain a physical state, one would still need to project the mean-field solution to the physical subspace. However, the inclusion of the Lagrange multiplier term allowed us to access magnetic phases on the mean-field level.



Interestingly, the sole fact that a given slave-particle state is not an element of the physical subspace does not imply that expectation values of physical observables are incorrect. In the following, we review two known examples from the existing literature in which correct results are obtained even if the constraints are relaxed completely. The first example has already been given in our original work,\cite{Ruegg:2010b} namely the non-interacting system at half-filling. For completeness, we review here some of the main arguments. Ignoring the constraint on the physical Hilbert space Eq.~\eqref{eq:Ai}, a ground state of Eq.~\eqref{eq:Hs} with $U=0$ at half-filling is readily found:
\begin{equation}
H_0|\Phi_0\rangle=E_0|\Phi_0\rangle,\quad |\Phi_0\rangle=|\{1/2\}\rangle_I\otimes|\phi_0\rangle_f.
\end{equation}
Here, $I_i^x|\{1/2\}\rangle_I=+1/2 |\{1/2\}\rangle_I$ for all $i$ and $|\phi_0\rangle_f$ is the half-filled Fermi sea obtained from Eq.~\eqref{eq:Hs} by replacing all operators $I^x_i$ with $1/2$ at $U=0$. This ground state $|\Phi_0\rangle$ fulfills
\begin{equation}
\langle A_i\rangle_{\Phi_0}=0,\quad A_i=I_i^z+1/2-(n_i-1)^2,
\end{equation}
on average; it is however {\it not} an element of the physical subspace because
\begin{equation}
\langle Q_i\rangle_{\Phi_0}=\langle A_i^2\rangle_{\Phi_0}=1/2\neq 0.
\end{equation}
Yet, it can be easily shown\cite{Ruegg:2010b} that all expectation values of physical observables are correctly obtained from $|\Phi_0\rangle$. Hence, the correct physics in the non-interacting limit at half filling is obtained despite completely relaxing the constraint. However, if needed, one can obtain the physical ground state by projection of $|\Phi_0\rangle$ and the result is a state which is a equal-weight superposition of all the states obtained from $|\Phi_0\rangle$ by a gauge transformation.\cite{Ruegg:2010b}  

Another example in which the projection onto the physical subspace can be ignored has been pointed out by Schir\'o and Fabrizio in Ref.~\onlinecite{Schiro:2011} and elaborated on by Baruselli and Fabrizio in Ref.~\onlinecite{Baruselli:2012}. These authors studied the single-impurity Anderson model as well as the half-filled Hubbard model in the infinite dimensional limit for which they derived the partition function in the slave-spin representation
\begin{equation}
Z=\left(\frac{1}{2}\right)^NZ_{S}, \quad Z_{S}={\rm Tr}\left(e^{-\beta H_S}\right).
\end{equation}
Here, the interacting slave-spin Hamiltonian is given by
\begin{equation}
H_S=-\frac{4t}{\sqrt{z}}\sum_{\langle i,j\rangle\sigma}\left(I_i^xI_j^xf_{i\sigma}^{\dag}f_{j\sigma}+{\rm h.c.}\right)+\frac{U}{4}\sum_i(1-2I_i^z),
\end{equation}
and the coordination number $z$ is taken to infinity at the end of the calculation. Remarkably, the projection to the physical subspace effectively drops out and enters the calculation of the partition function only through the prefactor $(1/2)^N$ ($N$=number of sites) which accounts for the additional degrees of freedom in $H_S$.

A general way to study if the projection to the physical subspace qualitatively changes the physics associated with a given variational (mean-field) ansatz is to investigate the role of the ``gauge fluctuations". These are (unphysical) fluctuations of the mean-field parameters which enter the theory as a result of the gauge redundancy introduced by the slave-particle representation. For the $Z_2$ slave-spin theory, the role of the gauge fluctuations has been discussed in Ref.~\onlinecite{Ruegg:2012}. It was also found that in the strongly correlated limit a fractionalized non-Fermi liquid phase with $Z_2$ topological order and potentially gapless charge and spin excitations can exist.


\bibliography{biblio}

\end{document}